\documentclass[12pt]{article}
\usepackage[utf8]{inputenc}

\usepackage[top=20truemm,bottom=30truemm,left=23truemm,right=27truemm]{geometry}
\renewcommand{\baselinestretch}{1.2}

\usepackage{amsmath, amsfonts}
\usepackage{breqn}
\usepackage{braket}
\usepackage{tikz-feynhand}
\usepackage{cite}
\usepackage{hyperref}
\hypersetup{
    colorlinks=true,
    linkcolor=blue,
    citecolor=blue
}

\numberwithin{equation}{section}

\newcommand{\paren}[1]{\left(#1\right)}
\newcommand{\Paren}[1]{\left[#1\right]}

\newcommand{\abs}[1]{\left|#1\right|}

\newcommand {\beq}{\begin{equation}}
\newcommand {\eeq}{\end{equation}}
\newcommand {\beqa}{\begin{eqnarray}}
\newcommand {\eeqa}{\end{eqnarray}}
\newcommand {\n}{\nonumber \\}

\newcommand {\del}{\partial}

\newcommand{\pvectwo}{\vec{p}^{\hspace{2pt} 2}}

\def\dm2{\delta m^{2}}

\def\deltatilde{\tilde{\delta}}
\def\phic{\phi_{c}}

\allowdisplaybreaks[1]

\begin{document}

\begin{titlepage}
\renewcommand{\thefootnote}{\fnsymbol{footnote}}
\begin{normalsize}
\begin{flushright}
\begin{tabular}{l}
November 2022
\end{tabular}
\end{flushright}
\end{normalsize}

~~\\

\vspace*{0cm}
    \begin{Large}
       \begin{center}
         {Exact renormalization group for wave functionals}
       \end{center}
    \end{Large}
\vspace{1cm}

\begin{center}
           Takaaki K{\sc uwahara}\footnote
            {
e-mail address :
kuwahara.takaaki.15@shizuoka.ac.jp},
          Gota T{\sc anaka}\footnote
            {
e-mail address :
tanaka.gota.14@cii.shizuoka.ac.jp},
           Asato T{\sc suchiya}\footnote
           {
e-mail address : tsuchiya.asato@shizuoka.ac.jp}
           {and}
           Kazushi Y{\sc amashiro}\footnote
            {
 e-mail address : yamashiro.kazushi.17@shizuoka.ac.jp}\\
      \vspace{1cm}

{\it Department of Physics, Shizuoka University}\\
                {\it 836 Ohya, Suruga-ku, Shizuoka 422-8529, Japan}\\
         \vspace{0.3cm}
 {\it Graduate School of Science and Technology, Shizuoka University}\\
               {\it 836 Ohya, Suruga-ku, Shizuoka 422-8529, Japan}

\end{center}

\hspace{5cm}

\setcounter{footnote}{0}

\begin{abstract}
\noindent
Motivated by the construction of continuum tensor networks for interacting field theories,
which are relevant in understanding the emergence of space-time
in the gauge/gravity correspondence,
we derive a non-perturbative functional differential equation for wave functionals
in scalar field theories from the exact renormalization group equation.
We check the validity of the equation using
the perturbation theory. We calculate
the wave functional up to the first-order perturbation and verify that it satisfies the equation.
\end{abstract}
\vfill
\end{titlepage}
\vfil\eject

\section{Introduction}
It is recognized that the
emergence of space-time is an essential feature
of quantum gravity.
Tensor network models such as
the MERA \cite{Vidal:2007hda,Swingle:2009bg}, the HaPPY code \cite{Pastawski:2015qua} and the random tensor network \cite{Hayden:2016cfa} give networks that
can be interpreted as discrete bulk space emerging from a boundary theory through quantum entanglement \cite{Ryu:2006bv}.
It should be crucial
to construct continuum tensor networks from field theories to obtain continuum emergent space-time.

Continuum tensor networks for free-field theories have been successfully constructed
based on the variational method. This is called
the cMERA \cite{Haegeman:2011uy,Nozaki:2012zj}, which is
a continuum counterpart of the MERA.
Interestingly, the relationship between the information metric
and geometry was studied in Ref.\cite{Nozaki:2012zj}.
It is, however, important to construct continuum tensor networks
non-perturbatively
for interacting theories from the point of view of the
gauge/gravity correspondence\cite{Maldacena:1997re}, because the strong coupling regime in
boundary theories corresponds to classical geometry with quantum fluctuations.
It seems non-trivial to construct trial functions
for the cMERA in interacting theories (see Refs.\cite{Fernandez-Melgarejo:2019sjo,Fernandez-Melgarejo:2020fzw,Fernandez-Melgarejo:2021mza} for an approach based on non-linear canonical
transformations).
We should remark here that there is
a non-perturbative construction of the cMERA based on the Weyl transformation, which is called the path-integral optimization \cite{Caputa:2017yrh}.

In the
MERA, which represents a wave function of a quantum many-body system,
the layers of the network are interpreted as representing the energy scale, which corresponds to the bulk direction in the gauge/gravity
correspondence.
Thus, constructing continuum tensor networks is considered to be equivalent to
obtaining the scale dependence
of a wave functional in a quantum field theory,
which should be determined by the renormalization group\footnote{For early work
on renormalization of quantum field theory from a wave functional approach, see Refs.\nobreak\cite{Symanzik:1981wd,Luscher:1985iu,Minic:1994ff}}.
In this paper, to obtain the scale dependence of the wave functional non-perturbatively,
we consider an approach based on the exact renormalization group.
Note here that
there are works on a perturbative construction
of continuum tensor networks for interacting theories
based on the renormalization group \cite{Cotler:2018ufx} and
the exact renormalization group approach to the continuum tensor networks
for O($N$) vector model at free fixed point \cite{Fliss:2016ifp}.

The exact renormalization group (ERG)
\cite{Wilson:1973jj,Wegner:1972ih}
(for a review of the ERG,
see \cite{Morris:1993qb, Morris:1998da,Aoki:2000wm,Bagnuls:2000ae,Polonyi:2001se,Gies:2006wv,Pawlowski:2005xe,Igarashi:2009tj,Rosten:2010vm,Dupuis:2020fhh}) serves as a powerful non-perturbative method for
quantum field theories along with lattice field theories.
The ERG gives a functional differential equation that describes the scale-dependence of the effective action.
In this paper, we derive a functional differential
equation obeyed by wave functionals of the ground states in scalar field theories
from the
ERG equation (the Polchinski equation \cite{Polchinski:1983gv}).
While our equation is a non-perturbative one,
we check the validity of our equation using the perturbation theory.
We calculate the wave functional
up to the first-order perturbation and verify
that it satisfies our equation.
While our motivation is to study a tensor network
description of emergent space, we expect
our findings to contribute to
developments
in the non-perturbative studies of quantum field theories.

This paper is organized as follows.
In Sect. 2, we derive an ERG equation for the wave functionals of the ground states
in scalar field theories
from the ERG equation (the Polchinski equation).
We solve the equation to obtain a solution taking the Gaussian form and
derive an ERG equation for the interaction part of the wave functionals.
In Sect. 3, we perform the perturbative expansion of the wave functional up to the first order. We develop a systematic method for the perturbative expansion of the wave
functional based on the path integral.
In Sect. 4, we check the validity of the ERG equation for the
wave functional derived in Sect. 2 by using
the results obtained in Sect. 3.
Section 5 is devoted to the conclusion and discussion.
In the appendix, the flow equation for the mass counterterm is derived.


\section{The ERG equation for wave functionals}
In this section, we derive the ERG equation
for the wave functionals of the ground states
in scalar field theories from the Polchinski equation.
Throughout this paper, we work in ($d+1$)-dimensional
Euclidean space-time, where the time direction is parameterized
by $\tau$, and use the following compact notations
for the integrals:
\begin{align}
   \int_p \equiv \int \frac{d^{d+1}p}{(2\pi)^{d+1}} \ , \quad
   \int_{\vec{p}}\equiv \int \frac{d^dp}{(2\pi)^d}
   \ ,
   \quad \int_{\tau} \equiv \int d\tau \ ,
\end{align}
where $p$ stands for the momentum in $d+1$ dimensions, while
$\vec{p}$ stands for the $d$-dimensional spatial part of $p$.
We also introduce a shorthand notation
\begin{align}
\tilde{\delta}(\vec{p}) = (2\pi)^d \delta(\vec{p})
\end{align}
and define $V$ by
\begin{align}
V=\tilde{\delta}(0)
\end{align}
which is the volume of space.
In what follows, we frequently use $\phi(p)$ and $\phi(\tau,\vec{p})$, which transform each other by
\begin{align}
\phi(p)=\phi(E,\vec{p}) =\int d\tau  \phi(\tau,\vec{p}) e^{-iE\tau} \ .
\end{align}

\subsection{The Polchinski equation}
First, we briefly review the Polchinski equation
for scalar field theories.
The ERG equations for scalar field theories have the following
general structure \cite{Latorre:2000qc,Arnone:2002yh,Arnone:2005fb,Morris:1999px}
\begin{align}
    - \Lambda \frac{\partial}{\partial \Lambda} e^{-S_\Lambda[\phi]}
    = \int_p \frac{\delta}{\delta \phi(p)} \left[
        G_\Lambda[\phi](p) e^{-S_\Lambda[\phi]} \right] ,
        \label{General ERG}
\end{align}
where $\Lambda$ is the effective cutoff and $S_{\Lambda}$
is the effective action at the scale $\Lambda$.
The functional $G_\Lambda[\phi](p)$, which also depends on $p$, is required to correspond to
a continuum blocking
procedure and to ensure the UV regularization of the equation.
The structure in Eq.(\ref{General ERG}) ensures the physical
requirement that
the partition function is unchanged under the
infinitesimal change of the effective cutoff $\Lambda$:
\begin{align}
    - \Lambda \partial_\Lambda Z
    = - \Lambda \partial_\Lambda \int \mathcal{D} \phi e^{-S_\Lambda[\phi]}
    = \int_p \mathcal{D} \phi \frac{\delta}{\delta \phi(p)} \left[
        G_\Lambda[\phi](p) e^{-S_\Lambda[\phi]} \right]
    = 0 \ .
\end{align}

Typically $G_\Lambda[\phi](p)$ takes the following form
\begin{align}
    G_\Lambda[\phi](p) = \frac{1}{2} \dot{C}_\Lambda(p)
        \frac{\delta}{\delta \phi(-p)} (S_\Lambda - 2 \hat{S}) \ ,
    \label{G}
\end{align}
where $\dot{C}_\Lambda \equiv - \Lambda \partial_\Lambda C_\Lambda $ is an ERG integration kernel
that incorporates the UV regularization and
specifies the coarse-graining procedure with
$\hat{S}$, which is called the seed action.
The Polchinski equation \cite{Polchinski:1983gv} corresponds to setting
the seed action $\hat{S}$ to $S_0$,
i.e., the free part of the effective action $S_{\Lambda}$ taking the form
\begin{align}
	S_0 = \int_p \frac{1}{2} \phi(p) C^{-1}_\Lambda(p) \phi(-p) \ .
	\label{S_0}
\end{align}
It is easily checked that
$S_0$ satisfies Eq.(\ref{General ERG}) with Eq.(\ref{G}).
Now Eq.(\ref{General ERG}) reduces to
\begin{align}
	- \Lambda \frac{\partial}{\partial \Lambda} e^{-S_\Lambda[\phi]}
		= \int_p \frac{\delta}{\delta \phi(p)} \left[ \frac{1}{2} \dot{C}_\Lambda(p)
		\left\{ \frac{\delta}{\delta \phi(-p)} (S_\Lambda - 2 S_0) \right\} e^{-S_\Lambda[\phi]} \right]  \ .
	\label{FRG for Effective Action}
\end{align}
By decomposing the effective action into the free part and
the interaction part as
\begin{align}
    S_\Lambda = S_0 + S_{\mathrm{int}}  \ ,
\end{align}
we obtain from Eq.(\ref{FRG for Effective Action})
a conventional form of
the Polchinski equation for $S_{\mathrm{int}}$:
\begin{align}
	- \Lambda \frac{\partial}{\partial \Lambda} e^{-S_{\mathrm{int}}}
		= - \frac{1}{2} \int_p \dot{C}_\Lambda(p) \frac{\delta^2}{\delta \phi(p) \delta \phi(-p)}
		e^{-S_{\mathrm{int}}}  \ .
	\label{Polchinski eq for Interaction}
\end{align}

\subsection{The ERG equation for wave functionals}
Next, we derive the ERG equation for the wave functional of the ground state
from the Polchinski equation (\ref{FRG for Effective Action}).

Note first that the ground-state wave functional $\Psi_\Lambda[\varphi]$ has the following path-integral
representation
\begin{align}
    \Psi_\Lambda[\varphi]
    = \int_{\phi(0,\vec{p}) = \varphi(\vec{p})} \mathcal{D} \phi
        e^{- \int_{- \infty}^0 d\tau L_\Lambda [\phi] }  \ ,
        \label{path integral representation}
\end{align}
where $L_{\Lambda}$ is the effective Lagrangian and
the boundary condition that the field $\phi(\tau,\vec{p})$
at $\tau=0$ is fixed to $\varphi(\vec{p})$ is imposed:
\begin{align}
\phi(0,\vec{p}) = \varphi(\vec{p}) \ .
\label{bc at tau=0}
\end{align}
We assume that $L_{\Lambda}$ is real so that
$\Psi_{\Lambda}[\varphi]$ is also real.
This implies that $\Psi_{\Lambda}[\varphi]$
is also represented as
\begin{align}
    \Psi_\Lambda[\varphi]
    = \int_{\phi(0,\vec{p}) = \varphi(\vec{p})} \mathcal{D} \phi
        e^{- \int_{0}^{\infty} d\tau L_\Lambda [\phi] }  \ .
        \label{another representation}
\end{align}

We see from Eqs.(\ref{path integral representation}) and
(\ref{another representation})
that the square of the wave functional $\Psi_\Lambda[\varphi]$
is represented in terms of the effective action
$S_{\Lambda}=\int_{-\infty}^{\infty}d\tau L_{\Lambda}$ as
\begin{align}
	\Psi_\Lambda^2[\varphi] = \int \mathcal{D} \phi \prod_{\vec{p}}
	\delta [\phi(0,\vec{p}) - \varphi(\vec{p})]
	e^{-S_\Lambda[\phi]}  \ .
	\label{square of Psi}
\end{align}
We act $-\Lambda\frac{\partial}{\partial \Lambda}$ on
both sides of (\ref{square of Psi}).
The left-hand side reduces to
\begin{align}
2 \Psi_\Lambda \left( - \Lambda \frac{\partial}{\partial \Lambda} \Psi_\Lambda \right) \ .
\label{LHS}
\end{align}
On the right-hand side, we use Eq.(\ref{FRG for Effective Action}) to obtain
\begin{align}
	&\int \mathcal{D} \phi \prod_{\vec{k}}
	    \delta [\phi(0,\vec{k}) - \varphi(\vec{k})]
		\int_p \frac{\delta}{\delta \phi(p)} \left[ \frac{1}{2} \dot{C}_\Lambda(p)
		\left\{ \frac{\delta}{\delta \phi(-p)} (S_\Lambda - 2 S_0) \right\} e^{-S_\Lambda} \right]  \ .
		\label{RHS_1}
\end{align}
Substituting Eq.(\ref{S_0}) into Eq.(\ref{RHS_1}) and switching
to the coordinate representation in the time direction
yields
\begin{align}
	 &\int \mathcal{D} \phi \prod_{\vec{k}}
	    \delta [\phi(0,\vec{k}) - \varphi(\vec{k})]
		\int_{\tau,\tau',\hspace{1pt} \vec{p}} \left[ - \frac{1}{2} \dot{C}_\Lambda(\tau-\tau',\vec{p})
			\frac{\delta^2}{\delta \phi(\tau,\vec{p}) \delta \phi(\tau',-\vec{p})} e^{-S_\Lambda[\phi]} \right]	\nonumber	\\
		&\hspace{10pt}- \int \mathcal{D} \phi \prod_{\vec{k}}
		\delta [\phi(0,\vec{k}) - \varphi(\vec{k})]
		\int_{\tau,\tau',\tau'', \hspace{1pt}\vec{p}} \frac{\delta}{\delta \phi(\tau,\vec{p})} \left[ \dot{C}_\Lambda(\tau-\tau',\vec{p})
			C^{-1}_\Lambda(\tau'-\tau'',\vec{p}) \phi(\tau'',\vec{p}) e^{-S_\Lambda[\phi]} \right] \ .
			\label{RHS_2}
\end{align}
By noting that the first line in Eq.(\ref{RHS_2})
is the integral of the total derivative except at $\tau=\tau'=0$,
as is the second line except at $\tau=0$,
we obtain
\begin{align}
	 &\int \mathcal{D} \phi \prod_{\vec{k}}
	    \delta [\phi(0,\vec{k}) - \varphi(\vec{k})]
		\int_{\tau,\tau',\hspace{1pt}\vec{p}} \left[ - \frac{1}{2} \dot{C}_\Lambda(0,\vec{p})
			\frac{\delta^2}{\delta \phi(0,\vec{p}) \delta \phi(0,-\vec{p})} e^{-S_\Lambda[\phi]} \right]	\nonumber	\\
		&\hspace{10pt}- \int \mathcal{D} \phi \prod_{\vec{k}}
		\delta [\phi(0,\vec{k}) - \varphi(\vec{k})]
		\int_{\tau',\tau'',\hspace{1pt}\vec{p}} \frac{\delta}{\delta \phi(0,\vec{p})} \left[ \dot{C}_\Lambda(-\tau',\vec{p})
			C^{-1}_\Lambda(\tau'-\tau'',\vec{p}) \phi(\tau'',\vec{p}) e^{-S_\Lambda[\phi]} \right] \ .
			\label{RHS_3}
\end{align}
Here we assume that $C_{\Lambda}(\tau,\vec{p})$
is factorized as
\begin{align}
    C_{\Lambda}(\tau,\vec{p})=f(\tau,\vec{p})g_{\Lambda}(\vec{p}) \ ,
    \label{factorization}
\end{align}
where $f$ is independent of $\Lambda$.
Namely, only the spatial components of the momentum have the UV cutoff.
This implies that
\begin{align}
   \int_{\tau''} \dot{C}_{\Lambda}(\tau-\tau'',\vec{p})C^{-1}_{\Lambda}(\tau''-\tau',\vec{p})
   = \delta(\tau-\tau') \frac{\dot{g}_{\Lambda}(\vec{p})}{g_{\Lambda}(\vec{p})}
   = \delta(\tau-\tau') \frac{\dot{C}_{\Lambda}(0,\vec{p})}{C_{\Lambda}(0,\vec{p})} \ .
   \label{CdotCinverse}
\end{align}
We substitute Eq.(\ref{CdotCinverse}) into Eq.(\ref{RHS_3}),
perform the partial integration with respect to
$\phi(0,\vec{p})$ and use the relation
\begin{align}
    \frac{\delta}{\delta \phi(0,\vec{p})} \prod_{\vec{k}}
        \delta [\phi(0, \vec{k}) - \varphi(\vec{k})]
    = - \frac{\delta}{\delta \varphi(\vec{p})} \prod_{\vec{k}}
        \delta [\phi(0, \vec{k}) - \varphi(\vec{k})]
\end{align}
to gain
\begin{align}
	& - \frac{1}{2} \int_{\vec{p}} \dot{C}_\Lambda(0,\vec{p})
			\frac{\delta^2}{\delta \varphi(\vec{p})
			\delta \varphi(-\vec{p})}
		\int \mathcal{D} \phi \prod_{\vec{k}}
		\delta [\phi(0,\vec{k}) - \varphi(\vec{k})]
		e^{-S_\Lambda[\phi]}	\nonumber	\\
		&\hspace{10pt}- \int_{\vec{p}}
		\frac{\delta}{\delta \varphi(\vec{p})}
		\left[ \frac{\dot{C}_{\Lambda}(0,\vec{p})}{C_{\Lambda}(0,\vec{p})}
		\varphi(\vec{p})  \int \mathcal{D} \phi
		\prod_{\vec{k}}\delta [\phi(0,\vec{k}) - \varphi(\vec{k})]
		e^{-S_\Lambda[\phi]} \right] \ .
		\label{RHS_4}
\end{align}
Furthermore, by using Eq.(\ref{square of Psi}),
we rewrite Eq.(\ref{RHS_4}) as
\begin{align}
	& - \Psi_\Lambda \int_{\vec{p}} \dot{C}_\Lambda(0,\vec{p})
		\left\{ \frac{\delta^2 \Psi_\Lambda}{\delta \varphi(\vec{p}) \delta \varphi(-\vec{p})}
			+ \frac{1}{\Psi_\Lambda} \frac{\delta \Psi_\Lambda}{ \delta \varphi(\vec{p})}
			\frac{\delta \Psi_\Lambda}{ \delta \varphi(-\vec{p})} \right\}
			\nonumber	\\
			&\hspace{10pt} -2 \Psi_\Lambda \int_{\vec{p}}
			    \frac{\dot{C}_{\Lambda}(0,\vec{p})}{C_{\Lambda}(0,\vec{p})}
			     \varphi(\vec{p})
				\frac{\delta \Psi_\Lambda}{\delta \varphi(\vec{p})}
			- \Psi_\Lambda^2 V \int_{\vec{p}} \frac{\dot{C}_{\Lambda}(0,\vec{p})}{C_{\Lambda}(0,\vec{p})} \ .
			\label{RHS_5}
\end{align}
Finally, combining Eqs.(\ref{LHS}) and (\ref{RHS_5})
leads us to the ERG equation for
the ground-state wave functional $\Psi_\Lambda [\varphi]$:
\begin{align}
	- \Lambda \frac{\partial}{\partial \Lambda} \Psi_\Lambda
	= &- \frac{1}{2} \int_{\vec{p}} \dot{C}_\Lambda(0,\vec{p})
		\left\{ \frac{\delta^2 \Psi_\Lambda}{\delta \varphi(\vec{p}) \delta \varphi(-\vec{p})}
			+ \frac{1}{\Psi_\Lambda} \frac{\delta \Psi_\Lambda}{ \delta \varphi(\vec{p})} \frac{\delta \Psi_\Lambda}{ \delta \varphi(-\vec{p})} \right\}
			\nonumber	\\
			&- \int_{\vec{p}}
			\frac{\dot{C}_{\Lambda}(0,\vec{p})}{C_{\Lambda}(0,\vec{p})}
			 \varphi(\vec{p})
			\frac{\delta \Psi_\Lambda}{\delta \varphi(\vec{p})}
			- \frac{V}{2} \Psi_\Lambda \int_{\vec{p}} \frac{\dot{C}_{\Lambda}(0,\vec{p})}{C_{\Lambda}(0,\vec{p})} \ .
	\label{FRG for Wave Functionals}
\end{align}

\subsection{A Gaussian solution}
As a demonstration of solving Eq.(\ref{FRG for Wave Functionals}), we search for a solution
that has the Gaussian form:
\begin{align}
    \Psi_0[\varphi]
    =\mathcal{N}_{\Lambda} \exp \left[-\frac{1}{2}
    \int_{\vec{p}} \varphi(\vec{p})
    \mathcal{M}_{\Lambda}(\vec{p})\varphi(-\vec{p})\right] \ .
    \label{Gaussian solution}
\end{align}
This solution would
correspond to the case in which $S_{\Lambda}=S_0$.
By substituting Eq.(\ref{Gaussian solution}) into
Eq.(\ref{FRG for Wave Functionals}),
we obtain flow equations for $\mathcal{N}_{\Lambda}$ and $\mathcal{M}_{\Lambda}(\vec{p})$:
\begin{align}
    -\Lambda\frac{\partial}{\partial \Lambda}
    \ln\mathcal{N}
    &=\frac{V}{2}\int_{\vec{p}}\dot{C}_{\Lambda}(0,\vec{p})(\mathcal{M}_{\Lambda}(\vec{p})
    -C^{-1}(0,\vec{p})) \ , \label{N} \\
     -\Lambda\frac{\partial}{\partial \Lambda}
    \mathcal{M}_{\Lambda}(\vec{p}) &=
    2\dot{C}_{\Lambda}(0,\vec{p})\mathcal{M}_{\Lambda}(\vec{p})(\mathcal{M}_{\Lambda}(\vec{p})
    -C^{-1}(0,\vec{p})) \ .
    \label{M}
\end{align}
We can find the general solution to Eq.(\ref{M}) as
\begin{align}
   \mathcal{M}_{\Lambda}(\vec{p})=\frac{1}{2C_{\Lambda}(0,\vec{p})+\alpha(\vec{p})C_{\Lambda}^2(0,\vec{p})} \ ,
   \label{general solution for M}
\end{align}
where $\alpha(\vec{p})$ is an arbitrary function
of $\vec{p}$.
Here we assume that $S_0$ reduces to the standard
free action in the $\Lambda\rightarrow\infty$
limit:
\begin{align}
S_0|_{\Lambda\rightarrow\infty}= \frac{1}{2}\int_p \phi(p) (p^2+m^2) \phi(-p) \ .
\label{standard S_0}
\end{align}
namely $C_{\infty}(p)=1/(p^2+m^2)$,
which implies $C_{\infty}(0,\vec{p})=1/(2\omega_{\vec{p}})$,
where $\omega_{\vec{p}}=\sqrt{\vec{p}^2+m^2}$.
 On the other hand, when the action is given
 by Eq.(\ref{standard S_0}), the ground-state wave functional is given by Eq.(\ref{Gaussian solution})
 with $\mathcal{M}_{\Lambda}(\vec{p})=\omega_{\vec{p}}$ \cite{Hatfield:1992rz}.
 Thus, we imposed the boundary condition
 $\mathcal{M}_{\infty}(\vec{p})=\omega_{\vec{p}}$,
 which fixes $\alpha(\vec{p})=0$ in Eq.(\ref{general solution for M}).
 Then, we can solve Eq.(\ref{N}) as
 \begin{align}
     \mathcal{N}_{\Lambda}=N_0
     \exp \left[ -\frac{V}{4}\int_{\vec{p}} \ln C_{\Lambda}(0,\vec{p}) \right] \ .
 \end{align}
 with $N_0$ being an arbitrary constant.
 We impose the boundary condition for $\mathcal{N}_{\Lambda}$ by requiring
 the normalization condition $1=\int \mathcal{D}\varphi |\Psi_0[\varphi]|^2$ in
 the $\Lambda\rightarrow\infty$ limit.
 This fixes $N_0$ to 1.
 We therefore obtain a Gaussian solution
 to Eq.(\ref{FRG for Wave Functionals}):
 \begin{align}
 \Psi_0[\varphi]
 =\exp \left[-\frac{1}{2}\int_{\vec{p}}
 \varphi(\vec{p})
 \frac{1}{2C_{\Lambda}(0,\vec{p})} \varphi(-\vec{p}) -\frac{V}{4}\int_{\vec{p}} \ln C_{\Lambda}(0,\vec{p})\right] .
 \label{Psi_0 solution}
 \end{align}
 In Sect. 3, we indeed obtain the above result by calculating
the path integral directly in the free case in which
$S_{\Lambda}=S_0$.

\subsection{The ERG equation for the interaction part of wave functionals}
We parameterize the ground-state wave functional as
\begin{align}
    \Psi_{\Lambda}[\varphi] = e^{I[\varphi]} \Psi_0[\varphi]
    \label{parameterization of Psi} \ .
\end{align}
$I[\varphi]$ is interpreted as representing the contribution of the interaction to the wave functional.
By substituting Eqs.(\ref{Psi_0 solution}) and (\ref{parameterization of Psi})
into Eq.(\ref{FRG for Wave Functionals}), we obtain the ERG equation for $I$:
\begin{align}
    - \Lambda \frac{\partial}{\partial \Lambda} I
    &= - \frac{1}{2}\int_{\vec{p}} \dot{C}(0,\vec{p})\left[
        \frac{\delta^2 I}{\delta \varphi(\vec{p}) \delta \varphi(-\vec{p})} +\frac{\delta I}{\delta\varphi(\vec{p})}\frac{\delta I}{\delta\varphi(-\vec{p})}\right] \ .
        \label{FRG for I}
\end{align}
This equation is a non-perturbative functional differential
equation for the interaction part of the ground-state wave functionals,
which is a counterpart of Eq.(\ref{Polchinski eq for Interaction}).

In the remaining part of this paper,
we use the following $C_{\Lambda}$:
\begin{align}
    C_{\Lambda}(p) =\frac{K(\pvectwo/\Lambda^2)}{p^2+m^2} \ , \;\;\;
    C_{\Lambda}(0,\vec{p})=\frac{K(\pvectwo/\Lambda^2)}{2\omega_{\vec{p}}}
    \label{C}
\end{align}
which has the property shown in
 (\ref{factorization}).
$K(x)$ is assumed to have the following properties:
$K(0)=1$, $K(x)\sim 1$ for $x<1$, and
$K(x)$ damps rapidly for $x>1$.
In this case, $\dot{C}_{\Lambda}(0,\vec{p})$ is
given by
\begin{align}
    \dot{C}_{\Lambda}(0,\vec{p})=\frac{\dot{K}(\pvectwo/\Lambda^2)}{2\omega_{\vec{p}}} \ ,
    \label{D}
\end{align}
where $\dot{K}(\pvectwo/\Lambda^2)=-\Lambda\partial_{\Lambda}K(\pvectwo/\Lambda^2)$.

\section{The perturbative wave functional}

In this section, we calculate the ground-state wave functional given by Eq.(\ref{path integral representation}) to the first-order perturbation. For technical convenience,
we introduce an infrared cutoff $T$ in the time direction and
eventually take the $T\rightarrow\infty$ limit:
  \begin{equation}
    \Psi_\Lambda[\varphi]  = \lim_{T \to \infty}
    \int_{\phi(0,\vec{p}) = \varphi(\vec{p})}
    \mathcal{D} \phi \ e^{-\int_{-T}^0 d\tau L_{\Lambda}} \ .
  \end{equation}
We can impose an arbitrary boundary condition at $\tau=-T$ in addition to Eq.(\ref{bc at tau=0}).
Here, for convenience, we impose the condition
\begin{align}
\phi(-T,\vec{p})=0 \ .
\label{bc at tau=-T}
\end{align}
The effective Lagrangian $L_{\Lambda}$ consists of
  the free part $L_0$ and the interaction part $L_{\mathrm{int}}$.
The free part $L_0$ is read off from $S_0$ in Eq.(\ref{S_0}) with Eq.(\ref{C}) as
  \begin{align}
    L_0 &= \int_{\vec{p}} \  \frac{1}{2} K_{\vec{p}}^{-1}
      \Paren{
        \partial_\tau \phi(\tau, \vec{p}) \partial_\tau \phi(\tau, -\vec{p})
        + \omega_{\vec{p}}^2 \phi(\tau,\vec{p}) \phi(\tau, -\vec{p})
      }\  \ ,
\end{align}
where we have introduced a shorthand notation
\begin{align}
K_{\vec{p}}=K(\vec{p}^{\hspace{2pt} 2}/\Lambda^2) \ .
\end{align}
We assume that the interaction part $L_{\mathrm{int}}$ consists of
the mass counterterm and the $\phi^4$ interaction term as
\begin{equation}
\begin{split}
      L_{\mathrm{int}} &= \frac{\dm2}{2} \int_{\vec{p}} \phi(\tau,\vec{p}) \phi(\tau, -\vec{p}) \\
        &\quad  + \frac{\lambda}{4!} \int_{\vec{p_1} \dots \vec{p_4}} \phi(\tau, \vec{p}_1) \phi(\tau, \vec{p}_2) \phi(\tau, \vec{p}_3) \phi(\tau, \vec{p}_4) \tilde{\delta}(\vec{p}_1 + \vec{p}_2 + \vec{p}_3 + \vec{p}_4)
        \ .
\label{Lint}
\end{split}
\end{equation}

In order to perform the perturbative expansion,
 we expand $\phi(\tau, \vec{p})$ around a classical solution  $\phi_c(\tau,\vec{p})$
in the free part $L_0$ as
  \begin{equation}
    \label{eq: expanding around the classical solution}
      \phi(\tau, \vec{p}) = \phic(\tau, \vec{p}) + \chi(\tau, \vec{p}) \ .
  \end{equation}
Namely, $\phi_c(\tau, \vec{p})$ satisfies the equation of motion
 \begin{equation}
    \label{eq: EoM}
    \partial_\tau^2 \phic(\tau, \vec{p}) = \omega_{\vec{p}}^2 \phic(\tau, \vec{p}) \ .
  \end{equation}
Here we set
\begin{align}
\phi_c(0, \vec{p}) = \varphi(\vec{p}) \ ,  \;\; \phi_c(-T, \vec{p}) = 0 \ , \;\;
\chi(0, \vec{p}) = 0 \ , \;\; \chi(-T, \vec{p}) = 0
\label{bc}
\end{align}
such
that the boundary conditions (\ref{bc at tau=0}) and (\ref{bc at tau=-T})
are satisfied.
We see from Eqs.(\ref{eq: EoM}) and (\ref{bc}) that $\phi(\tau,\vec{p})$ is explicitly given by
  \begin{align}
      \phic(\tau,\vec{p})
      &= \frac{e^{\omega_{\vec{p}} (\tau + T)} - e^{-\omega_{\vec{p}}(\tau + T)}}{e^{\omega_{\vec{p}} T} - e^{-\omega_{\vec{p}} T}} \varphi(\vec{p}) \label{phi_c}
\nonumber\\
      &\xrightarrow{T \to \infty} \  e^{\omega_{\vec{p}} \tau} \varphi(\vec{p})
\end{align}
and $\chi(\tau,\vec{p})$ has the following Fourier expansion
\begin{align}
\chi(\tau,\vec{p}) = \sum_{n=1}^\infty  \sin \paren{\frac{n \pi}{T} \tau} \chi_n (\vec{p}) \ .
\label{chi}
\end{align}
Substituting Eq.(\ref{eq: expanding around the classical solution}) into $L_0$ and using Eqs.(\ref{eq: EoM}) and (\ref{bc}) gives rise to
\begin{align}
\int_{-T}^0 d\tau L_0= \frac{1}{2}\int_{\vec{p}} K_{\vec{p}}^{-1}\phi_c(\tau,\vec{p})\partial_{\tau}\phi_c(\tau,\vec{p})|_{\tau=0}
 \ + \ \frac{1}{2}\int_{-T}^0d\tau\int_{\vec{p}}K_{\vec{p}}^{-1}\chi(\tau,-\vec{p})(-\partial_{\tau}^2+\omega_{\vec{p}}^2)\chi(\tau,\vec{p}) \ .
\label{on-shell action}
\end{align}

We parameterize the ground-state wave functional as
Eq.(\ref{parameterization of Psi}).
First, noting that the lowest-order of the wave functional is given by
\begin{align}
\Psi_0[\varphi] = \lim_{T \to \infty} \int \mathcal{D}\chi e^{- \int_{-T}^0 d\tau L_0} \ ,
\end{align}
we see from (\ref{phi_c}) and the first term in (\ref{on-shell action}) that
  \begin{equation}
    \Psi_0[\varphi] = N_0\exp\Paren{-\int_{\vec{p}}\frac{1}{2}K_{\vec{p}}^{-1} \omega_{\vec{p}} \varphi(\vec{p})\varphi(-\vec{p})
} \ ,
   \label{Psi_0 from path integral}
  \end{equation}
where $N_0$ is the normalization constant which is fixed by the condition
$1 = \int \mathcal{D}\varphi |\Psi_0[\varphi]|^2$ as
\begin{align}
N_0=\exp \left[\frac{V}{4} \int_{\vec{p}} \log (2 K_{\vec{p}}^{-1} \omega_{\vec{p}}  ) \right] \ .
 \label{normalization factor}
\end{align}
Equation (\ref{Psi_0 from path integral}) with Eq.(\ref{normalization factor})
indeed agrees with Eq.(\ref{Psi_0 solution}).


  Next, we calculate the contribution of the interaction
  to the wave functional, $I[\varphi]$, perturbatively.
For that, we can use the Feynman diagrams in which the external and internal lines represent
$\varphi$ and the propagator for $\chi$, respectively.
We see from Eqs.(\ref{chi}) and (\ref{on-shell action}) that the propagator for $\chi$ is
given by
  \begin{align}
    \label{chi chi propagator}
      \braket{\chi(\tau,\vec{p})\chi(\tau',\vec{p}')}
      &= \sum_n \sin \paren{ \frac{n\pi}{T}\tau } \sin \paren{ \frac{n\pi}{T}\tau' }
        \frac{2}{T} \frac{K_{\vec{p}} \deltatilde(\vec{p}+\vec{p}')}{\omega_{\vec{p}}^2 + (\frac{n \pi}{T})^2} \nonumber \\
      &\xrightarrow{T \to \infty}
        \quad \frac{K_{\vec{p}}}{2 \omega_{\vec{p}}} \paren{e^{- \omega_{\vec{p}} \abs{\tau - \tau'}} - e^{\omega_{\vec{p}} (\tau + \tau')}} \deltatilde(\vec{p}+\vec{p}')
      \ .
  \end{align}
By substituting Eqs.(\ref{eq: expanding around the classical solution})
and (\ref{phi_c})
into $L_{\mathrm{int}}$, we can read off the interaction vertices for $\varphi$ and
$\chi$.
Here we represent the interaction with $\delta m^2$ by
the black dot vertex and the interaction  with $\lambda$ by
the plain vertex.
 $I[\varphi ]$ consists of the sum of the connected diagrams except the bubble diagrams and
a constant that is fixed by the normalization condition $1=\int \mathcal{D}\varphi
|\Psi[\varphi]|^2$.
Here we concentrate on the first-order perturbation.
The connected diagrams that we need are given by
  \begin{align}
    \begin{split}
        \begin{tikzpicture}[baseline=(a.base)]
          \begin{feynhand}
            \vertex (a) at (-1,0); \vertex [dot] (b) at (0,0) {}; \vertex (c) at (1,0);
            \propag[plain] (a) to (b);
            \propag[plain] (b) to (c);
          \end{feynhand}
        \end{tikzpicture}
        = - \frac{\dm2}{2} \int_{\vec{p}} \varphi(\vec{p}) \varphi(-\vec{p}) \frac{1}{2\omega_{\vec{p}}} \ ,
    \end{split}
    \\
    \nonumber \\
    \begin{split}
      \begin{tikzpicture}[baseline=0.3cm]
        \begin{feynhand}
          \vertex (a) at (-1,0); \vertex (b) at (0,0); \vertex (c) at (1,0);
          \vertex (d) at (0,1);
          \propag[plain] (a) to (b);
          \propag[plain] (b) to (c);
          \propag[plain] (b) to [half left, looseness=1.5] (d);
          \propag[plain] (b) to [half right, looseness=1.5] (d);
        \end{feynhand}
      \end{tikzpicture}
      = - \frac{\lambda}{4!} \int_{\vec{p}_1 \vec{p}_2} \varphi(\vec{p}_1) \varphi(-\vec{p}_1)
        \ \frac{3K_2}{2\omega_1 (\omega_1 + \omega_2)} \ ,
    \end{split}
    \\
    \nonumber \\
    \begin{split}
        \begin{tikzpicture}[baseline=0.35cm]
          \begin{feynhand}
            \vertex (a) at (0,1); \vertex (b) at (1,1);
            \vertex (c) at (0,0); \vertex (d) at (1,0);
            \propag[plain] (a) to (d);
            \propag[plain] (b) to (c);
          \end{feynhand}
        \end{tikzpicture}
        = -\frac{\lambda}{4!} \int_{\vec{p}_1 \cdots \vec{p}_4} \varphi_1 \cdots \varphi_4
          \ \frac{\deltatilde(\vec{p}_1 + \vec{p}_2 + \vec{p}_3 + \vec{p}_4)}{\omega_1 + \omega_2 + \omega_3 + \omega_4} \ ,
    \end{split}
    \label{diagrams}
  \end{align}
where we have introduced shorthand notations:
  $\varphi(\vec{p}_i) = \varphi_i$,
  $K(\vec{p}_i^{\hspace*{1mm} 2} / \Lambda^2) = K_i$, and $\omega_{\vec{p}_i} = \omega_i$.

After fixing the constant in $I[\varphi ]$ using the normalization condition,
we obtain the final result for the ground-state wave functional to the first-order perturbation as
  \begin{align}
  \label{eq: perturbative wave functional}
      \Psi_\Lambda[\varphi] =&  e^{I[\varphi ]} \Psi_0[\varphi]  \ , \\
      I[\varphi ] =& - \frac{\dm2}{2} \int_{\vec{p}} \varphi(\vec{p}) \varphi(-\vec{p}) \frac{1}{2\omega_{\vec{p}}} \nonumber \\
          &  - \frac{\lambda}{4!} \int_{\vec{p}_1 \vec{p}_2} \varphi(\vec{p}_1) \varphi(-\vec{p}_1) \ \frac{3K_2}{2\omega_1 (\omega_1 + \omega_2)} \nonumber  \\
          &   -\frac{\lambda}{4!} \int_{\vec{p}_1 \cdots \vec{p}_4} \varphi_1 \cdots \varphi_4
              \ \frac{\deltatilde(\vec{p}_1 + \vec{p}_2 + \vec{p}_3 + \vec{p}_4)}{\omega_1 + \omega_2 + \omega_3 + \omega_4}
      + \mathcal{C}
      %
  \end{align}
with
\begin{equation}
\mathcal{C}
 = \left\{ \frac{\delta m^2}{2}
        + \frac{\lambda}{4!} \int_{\vec{p}} \frac{6 K_{\vec{p}}}{2\omega_{\vec{p}}} \right\}
        \int_{\vec{k}} \frac{ K_{\vec{k}} V}{4 \omega_{\vec{k}}^2}-  \frac{\lambda}{4!} \int_{\vec{p}_1, \vec{p}_2} \frac{3 K_1 K_2 V}{\omega_1 \omega_2 (\omega_1 + \omega_2)} \ .
\label{calC}
\end{equation}
 The above result agrees with the one obtained using the canonical formalism \cite{Hatfield:1992rz}.
The formulation of the perturbative expansion  that we have developed here based on
the path integral can be applied straightforwardly and systematically to the higher-order perturbations.

\section{A perturbative check}
In this section, as a check of the validity of the ERG equation (\ref{FRG for Wave Functionals}) for the ground-state wave functional,
we show that the ground-state wave functional (\ref{eq: perturbative wave functional})
satisfies the ERG equation (\ref{FRG for Wave Functionals}).
Since the analysis in Sect. 2.3 ensures that
$\Psi_0$ in Eq.(\ref{eq: perturbative wave functional})
satisfies the ERG equation
(\ref{FRG for Wave Functionals}) in Sect. 2.3, our task is to
check that
$I[\varphi]$ in Eq.(\ref{eq: perturbative wave functional}) satisfies
Eq.(\ref{FRG for I}).

Equation (\ref{FRG for I}) to the first order perturbation reads
\begin{align}
    - \Lambda \frac{\partial}{\partial \Lambda} I
    &= - \int_{\vec{p}} \frac{\dot{K}_{\vec{p}}}{4 \omega_{\vec{p}}}
        \frac{\delta^2 I}{\delta \varphi(\vec{p}) \delta \varphi(-\vec{p})}  \ .
        \label{FRG for I 2}
\end{align}
Substituting $I$ in  Eq.(\ref{eq: perturbative wave functional}) into the left-hand side of Eq.(\ref{FRG for I 2}) yields
\begin{align}
    - \Lambda \frac{\partial}{\partial \Lambda} I
    = &- \left\{ \frac{\dot{\delta m^2}}{2}
        + \frac{\lambda}{4!} \int_{\vec{p}} \frac{6 \dot{K}_{\vec{p}}}{2\omega_{\vec{p}}} \right\}
        \int_{\vec{k}} \left[ \frac{1}{2\omega_{\vec{k}}} \varphi(\vec{k}) \varphi(-\vec{k})
            - \frac{K_{\vec{k}} V}{4 \omega_{\vec{k}}^2}\right]   \nonumber   \\
    &+  \left\{ \frac{\delta m^2}{2}
        + \frac{\lambda}{4!} \int_{\vec{p}} \frac{6 K_{\vec{p}}}{2\omega_{\vec{p}}} \right\}
        \int_{\vec{k}} \frac{ \dot{K}_{\vec{k}} V}{4 \omega_{\vec{k}}^2}   \nonumber   \\
    &+ \frac{\lambda}{4!} \int_{\vec{k}_i} \frac{3}{\omega_1 + \omega_2}
        \frac{\dot{K}_2}{2 \omega_2} \phi(\vec{k}_1) \phi(-\vec{k}_1)   \nonumber   \\
    &+ \frac{\lambda}{4!} \int_{\vec{k}_i} \frac{3}{\omega_1 + \omega_2}
        \frac{\dot{K}_2}{2 \omega_2}
        \left( - \frac{\dot{K}_1 V}{2 \omega_1} \right) \ ,
    \label{LHS of ERG for I}
\end{align}
while the right-hand side gives
\begin{align}
    - \int_{\vec{p}} \frac{\dot{K}_{\vec{p}}}{4 \omega_{\vec{p}}}
        \frac{\delta^2 I}{\delta \varphi(\vec{p}) \delta \varphi(-\vec{p})}
    = &\left\{ \frac{\delta m^2}{2}
        + \frac{\lambda}{4!} \int_{\vec{p}} \frac{6 K_{\vec{p}}}{2\omega_{\vec{p}}} \right\}
        \int_{\vec{k}} \frac{\dot{K}_{\vec{k}} V}{4\omega_{\vec{k}}^2}   \nonumber   \\
    &+ \frac{\lambda}{4!} \int_{\vec{k}_i} \frac{3}{\omega_{1} + \omega_{2}}
        \frac{\dot{K}_{2}}{2 \omega_{2}} \phi(\vec{k}_1) \phi(-\vec{k}_1)   \nonumber   \\
    &+ \frac{\lambda}{4!} \int_{\vec{k}_i} \frac{3}{\omega_{1} + \omega_{2}}
        \frac{\dot{K}_{2}}{2 \omega_{2}}
        \left( - \frac{\dot{K}_{1} V}{2 \omega_{1}} \right) \ .
    \label{RHS of ERG for I}
\end{align}
By using the flow equation for $\delta m^2$
\begin{align}
    \frac{ \dot{\delta m^2}}{2}
        = - \frac{\lambda}{4!} \int_{\vec{p}} \frac{6 \dot{K}_{\vec{p}}}{2\omega_{\vec{p}}}
    \label{Scale Dependence of Mass Counterterm}
\end{align}
shown in the appendix,
we can
see that Eq.(\ref{LHS of ERG for I}) agrees with Eq.(\ref{RHS of ERG for I}).
Therefore $\Psi_\Lambda = e^I \Psi_0$ in Eq.(\ref{eq: perturbative wave functional})
satisfies the ERG equation Eq.(\ref{FRG for Wave Functionals}).

\section{Conclusion and discussion}
In this paper, we derived the ERG equation for the ground
state wave functionals in scalar field theories from
the ERG equation (the Polchinski equation). We checked the validity
of the equation by performing a perturbative expansion up to
the first order. Here we developed a systematic method for the
perturbative expansion of the wave functionals.

While we considered the Polchinski equation in this paper,
we could have started with other ERG equations.
For instance, we can set the seed action $\hat{S}$ in (\ref{G}) to $S_{\Lambda}$.
Then, the ERG equation for the wave functional reads
\begin{align}
-\Lambda \frac{\partial}{\partial \Lambda} \Psi_{\Lambda}
=  \frac{1}{2} \int_{\vec{p}} \dot{C}_\Lambda(0,\vec{p})
		\left\{ \frac{\delta^2 \Psi_\Lambda}{\delta \varphi(\vec{p}) \delta \varphi(-\vec{p})}
			+ \frac{1}{\Psi_\Lambda} \frac{\delta \Psi_\Lambda}{ \delta \varphi(\vec{p})} \frac{\delta \Psi_\Lambda}{ \delta \varphi(-\vec{p})} \right\} \ .
\end{align}

As discussed in the introduction,
the wave functionals that are solutions to Eqs.(\ref{FRG for Wave Functionals}) or (\ref{FRG for I})
with Eq.(\ref{parameterization of Psi}) are considered to
represent continuum tensor networks.
By using the solutions, we can
calculate the Fisher information
metric which measures the distance between the states at different energy scales.
As discussed in Refs.\cite{Nozaki:2012zj,Fernandez-Melgarejo:2021mza},
we expect to extract dual bulk geometry from the Fisher information metric.
It is also desirable to obtain the dependence of entanglement entropy on the energy scale
by using the solutions\footnote{For the calculation of entanglement entropy
in interacting field theories, see \cite{Hertzberg:2012mn,Chen:2020ild,Metlitski:2009iyg,Akers:2015bgh,Cotler:2015zda,Fernandez-Melgarejo:2020utg,Whitsitt:2016irx,Hampapura:2018uho,Buividovich:2008kq,Itou:2015cyu,Rabenstein:2018bri,Iso:2021dlj}}.

We would like to develop a method to solve
Eqs.(\ref{FRG for Wave Functionals}) or (\ref{FRG for I}) based on an approximation such
as the derivative expansion or
a numerical method, for instance, a wave functional analog of a method proposed in
Ref.\cite{Cotler:2022fze}.

In this paper, we restricted ourselves to scalar field theories. Extension to gauge field theories is of course important (for a recent development in the ERG equation
for the effective action in gauge field theories, see Ref.\cite{Sonoda:2020vut}).

We hope to report progress in the above-mentioned issues in the near future.

\section*{Acknowledgments}
We would like to thank  S. Iso and T. Takayanagi for discussions.
A.T. was supported in part by JSPS KAKENHI Grant Numbers JP18K03614 and JP21K03532 and
 by MEXT KAKENHI Grant in Aid for Transformative
Research Areas A ``Extreme Universe” No. 22H05253.


\section*{Appendix: Flow equation for $\delta m^2$}
\renewcommand{\theequation}{A.\arabic{equation}}
\setcounter{equation}{0}
In this appendix, we derive the flow equation for $\delta m^2$
from Eq.(\ref{Polchinski eq for Interaction})
up to the first-order perturbation. By using Eq.(\ref{C}),
Eq.(\ref{Polchinski eq for Interaction}) is rewritten as
\begin{align}
  -\Lambda \frac{\del S_{\mathrm{int}}}{\del \Lambda}
  = - \frac{1}{2} \int_p \frac{\dot{K}_{\vec{p}}}{2\omega_{\vec{p}}}
  \left( \frac{\delta^2 S_{\mathrm{int}}}{ \delta \phi(p) \delta \phi(-p) } - \frac{\delta S_{\mathrm{int}}}{ \delta \phi(p) } \frac{\delta S_{\mathrm{int}}}{ \delta \phi(-p) } \right) \ .
  \label{polchinski eq}
\end{align}
$S_{\mathrm{int}}$ can be read off from Eq.(\ref{Lint}) as
\begin{align}
  S_{\mathrm{int}} = \frac{\delta m^2}{2}  \int_k \phi(k) \phi(-k)
      + \frac{\lambda}{4!}  \int_{k_i} (2\pi)^{d+1}\delta (\Sigma_{i=1}^4 k_i) \ \prod_{i=1}^4 \phi(k_i) \ .
      \label{action of interaction}
\end{align}
We substitute Eq.(\ref{action of interaction}) into Eq.(\ref{polchinski eq}) and focus on the $\phi^2$ terms.
The left-hand side of Eq.(\ref{polchinski eq}) gives
\begin{align}
  \frac{\delta \dot{m}^2}{2} \int_{k} \phi(k) \phi(-k)
  \label{lhs deltam2}
\end{align}
The first term on the right-hand side of Eq.(\ref{polchinski eq})
gives
\begin{align}
  & - \frac{1}{2} \int_p \frac{\dot{K}_{\vec{p}}}{2\omega_{\vec{p}}}\frac{\delta^2 }{ \delta \phi(p) \delta \phi(-p) }\left\{ \frac{\lambda}{4!}  \int_{k_i} (2\pi)^{d+1}\delta (\Sigma_{i=1}^4 k_i) \ \prod_{i=1}^4 \phi(k_i) \right\} \n
  & = - \frac{\lambda}{4} \int_{k} \frac{\dot{K}_{\vec{p}}}{2\omega_{\vec{k}}}\phi(k)\phi(-k)
  \label{rhs deltam2}
\end{align}
The second-term of Eq.(\ref{polchinski eq}) gives a term
that is higher order in $\lambda$. Thus, from Eqs.(\ref{lhs deltam2})
and (\ref{rhs deltam2}), we obtain the flow equation for
$\delta m^2$ up to the first order in $\lambda$ as
\begin{align}
  \frac{\dot{\delta m^2}}{2}
  = - \frac{\lambda}{4!} \int_{\vec{p}}
  \frac{6\dot{K}_{\vec{p}}}{ 2\omega_{\vec{p}}}  \ .
\end{align}


\begin{thebibliography}{9}
\renewcommand{\baselinestretch}{0.8}


\bibitem{Vidal:2007hda}
G.~Vidal,
Phys. Rev. Lett. \textbf{99}, no.22, 220405 (2007)
[arXiv:cond-mat/0512165 [cond-mat]].

\bibitem{Swingle:2009bg}
B.~Swingle,
Phys. Rev. D \textbf{86}, 065007 (2012)
[arXiv:0905.1317 [cond-mat.str-el]].

\bibitem{Pastawski:2015qua}
F.~Pastawski, B.~Yoshida, D.~Harlow and J.~Preskill,
JHEP \textbf{06}, 149 (2015)
[arXiv:1503.06237 [hep-th]].

\bibitem{Hayden:2016cfa}
P.~Hayden, S.~Nezami, X.~L.~Qi, N.~Thomas, M.~Walter and Z.~Yang,
JHEP \textbf{11}, 009 (2016)
[arXiv:1601.01694 [hep-th]].


\bibitem{Ryu:2006bv}
  S.~Ryu and T.~Takayanagi,
  Phys.\ Rev.\ Lett.\  {\bf 96}, 181602 (2006)
  [hep-th/0603001].

\bibitem{Haegeman:2011uy}
J.~Haegeman, T.~J.~Osborne, H.~Verschelde and F.~Verstraete,
Phys. Rev. Lett. \textbf{110}, no.10, 100402 (2013)
[arXiv:1102.5524 [hep-th]].

\bibitem{Nozaki:2012zj}
M.~Nozaki, S.~Ryu and T.~Takayanagi,
JHEP \textbf{10}, 193 (2012)
[arXiv:1208.3469 [hep-th]].

\bibitem{Maldacena:1997re}
J.~M.~Maldacena,
Adv. Theor. Math. Phys. \textbf{2}, 231-252 (1998)
[arXiv:hep-th/9711200 [hep-th]].

\bibitem{Fernandez-Melgarejo:2019sjo}
J.~J.~Fernandez-Melgarejo, J.~Molina-Vilaplana and E.~Torrente-Lujan,
Phys. Rev. D \textbf{100}, no.6, 065025 (2019)
[arXiv:1904.07241 [hep-th]].


\bibitem{Fernandez-Melgarejo:2020fzw}
J.~J.~Fernandez-Melgarejo and J.~Molina-Vilaplana,
JHEP \textbf{07}, 149 (2020)
[arXiv:2003.08438 [hep-th]].

\bibitem{Fernandez-Melgarejo:2021mza}
J.~J.~Fernandez-Melgarejo and J.~Molina-Vilaplana,
JHEP \textbf{04}, 020 (2022)
[arXiv:2107.13248 [hep-th]].


\bibitem{Caputa:2017yrh}
P.~Caputa, N.~Kundu, M.~Miyaji, T.~Takayanagi and K.~Watanabe,
JHEP \textbf{11}, 097 (2017)
[arXiv:1706.07056 [hep-th]].

\bibitem{Symanzik:1981wd}
K.~Symanzik,
Nucl. Phys. B \textbf{190}, 1-44 (1981).

\bibitem{Luscher:1985iu}
M.~Luscher,
Nucl. Phys. B \textbf{254}, 52-57 (1985).

\bibitem{Minic:1994ff}
D.~Minic and V.~P.~Nair,
Int. J. Mod. Phys. A \textbf{11}, 2749-2764 (1996)
[arXiv:hep-th/9406074 [hep-th]].


\bibitem{Cotler:2018ufx}
J.~Cotler, M.~R.~Mohammadi Mozaffar, A.~Mollabashi and A.~Naseh,
Fortsch. Phys. \textbf{67}, no.10, 1900038 (2019)
[arXiv:1806.02831 [hep-th]].



\bibitem{Fliss:2016ifp}
J.~R.~Fliss, R.~G.~Leigh and O.~Parrikar,
Phys. Rev. D \textbf{95}, no.12, 126001 (2017)
[arXiv:1609.03493 [hep-th]].

\bibitem{Wilson:1973jj}
K.~G.~Wilson and J.~B.~Kogut,
Phys. Rept. \textbf{12}, 75-199 (1974).

\bibitem{Wegner:1972ih}
F.~J.~Wegner and A.~Houghton,
Phys. Rev. A \textbf{8}, 401-412 (1973).

\bibitem{Morris:1993qb}
T.~R.~Morris,
Int. J. Mod. Phys. A \textbf{9}, 2411-2450 (1994)
[arXiv:hep-ph/9308265 [hep-ph]].

 \bibitem{Morris:1998da}
 T.~R.~Morris,
 Prog. Theor. Phys. Suppl. \textbf{131}, 395-414 (1998)
 [arXiv:hep-th/9802039 [hep-th]].

 \bibitem{Aoki:2000wm}
 K.~Aoki,
 Int. J. Mod. Phys. B \textbf{14}, 1249-1326 (2000)

 \bibitem{Bagnuls:2000ae}
 C.~Bagnuls and C.~Bervillier,
 Phys. Rept. \textbf{348}, 91 (2001)
 [arXiv:hep-th/0002034 [hep-th]].

 \bibitem{Polonyi:2001se}
 J.~Polonyi,
 Central Eur. J. Phys. \textbf{1}, 1-71 (2003)
 [arXiv:hep-th/0110026 [hep-th]].

 \bibitem{Gies:2006wv}
 H.~Gies,
 Lect. Notes Phys. \textbf{852}, 287-348 (2012)
 [arXiv:hep-ph/0611146 [hep-ph]].

 \bibitem{Pawlowski:2005xe}
 J.~M.~Pawlowski,
 Annals Phys. \textbf{322}, 2831-2915 (2007)
 [arXiv:hep-th/0512261 [hep-th]].

 \bibitem{Igarashi:2009tj}
 Y.~Igarashi, K.~Itoh and H.~Sonoda,
 Prog. Theor. Phys. Suppl. \textbf{181}, 1-166 (2010)
 [arXiv:0909.0327 [hep-th]].

\bibitem{Rosten:2010vm}
O.~J.~Rosten,
Phys. Rept. \textbf{511}, 177-272 (2012)
[arXiv:1003.1366 [hep-th]].

 \bibitem{Dupuis:2020fhh}
 N.~Dupuis, L.~Canet, A.~Eichhorn, W.~Metzner, J.~M.~Pawlowski, M.~Tissier and N.~Wschebor,
 Phys. Rept. \textbf{910}, 1-114 (2021)
 [arXiv:2006.04853 [cond-mat.stat-mech]].

\bibitem{Polchinski:1983gv}
J.~Polchinski,
Nucl. Phys. B \textbf{231}, 269-295 (1984).


\bibitem{Latorre:2000qc}
J.~I.~Latorre and T.~R.~Morris,
JHEP \textbf{11}, 004 (2000)
[arXiv:hep-th/0008123 [hep-th]].


\bibitem{Arnone:2002yh}
S.~Arnone, A.~Gatti and T.~R.~Morris,
JHEP \textbf{05}, 059 (2002)
[arXiv:hep-th/0201237 [hep-th]].

%
%

%
%
%
%
%
%
%
%
%
%
%
%
%
%
%
%

\bibitem{Arnone:2005fb}
S.~Arnone, T.~R.~Morris and O.~J.~Rosten,
Eur. Phys. J. C \textbf{50}, 467-504 (2007)
[arXiv:hep-th/0507154 [hep-th]].

\bibitem{Morris:1999px}
T.~R.~Morris,
Nucl. Phys. B \textbf{573}, 97-126 (2000)
[arXiv:hep-th/9910058 [hep-th]].


\bibitem{Hatfield:1992rz}
B.~Hatfield,
``Quantum field theory of point particles and strings,''
Perseus (1998).










\bibitem{Buividovich:2008kq}
P.~V.~Buividovich and M.~I.~Polikarpov,
Nucl. Phys. B \textbf{802}, 458-474 (2008)
[arXiv:0802.4247 [hep-lat]].

\bibitem{Metlitski:2009iyg}
M.~A.~Metlitski, C.~A.~Fuertes and S.~Sachdev,
Phys. Rev. B \textbf{80}, no.11, 115122 (2009)
[arXiv:0904.4477 [cond-mat.stat-mech]].

\bibitem{Hertzberg:2012mn}
M.~P.~Hertzberg,
J. Phys. A \textbf{46}, 015402 (2013)
[arXiv:1209.4646 [hep-th]].

\bibitem{Cotler:2015zda}
J.~Cotler and M.~T.~Mueller,
Annals Phys. \textbf{365}, 91-117 (2016)
[arXiv:1509.05685 [hep-th]].

\bibitem{Akers:2015bgh}
C.~Akers, O.~Ben-Ami, V.~Rosenhaus, M.~Smolkin and S.~Yankielowicz,
JHEP \textbf{03}, 002 (2016)
[arXiv:1512.00791 [hep-th]].

\bibitem{Itou:2015cyu}
E.~Itou, K.~Nagata, Y.~Nakagawa, A.~Nakamura and V.~I.~Zakharov,
PTEP \textbf{2016}, no.6, 061B01 (2016)
[arXiv:1512.01334 [hep-th]].

\bibitem{Whitsitt:2016irx}
S.~Whitsitt, W.~Witczak-Krempa and S.~Sachdev,
Phys. Rev. B \textbf{95}, no.4, 045148 (2017)
[arXiv:1610.06568 [cond-mat.str-el]].

\bibitem{Hampapura:2018uho}
H.~R.~Hampapura, A.~Lawrence and S.~Stanojevic,
Phys. Rev. B \textbf{100}, no.13, 134412 (2019)
[arXiv:1811.04109 [hep-th]].

\bibitem{Rabenstein:2018bri}
A.~Rabenstein, N.~Bodendorfer, P.~Buividovich and A.~Sch\"afer,
Phys. Rev. D \textbf{100}, no.3, 034504 (2019)
[arXiv:1812.04279 [hep-lat]].

\bibitem{Chen:2020ild}
Y.~Chen, L.~Hackl, R.~Kunjwal, H.~Moradi, Y.~K.~Yazdi and M.~Zilh\~ao,
JHEP \textbf{11}, 114 (2020)
[arXiv:2002.00966 [hep-th]].

\bibitem{Fernandez-Melgarejo:2020utg}
J.~J.~Fernandez-Melgarejo and J.~Molina-Vilaplana,
JHEP \textbf{02}, 106 (2021)
[arXiv:2010.05574 [hep-th]].

\bibitem{Iso:2021dlj}
S.~Iso, T.~Mori and K.~Sakai,
Symmetry \textbf{13}, no.7, 1221 (2021)
[arXiv:2105.14834 [hep-th]].

\bibitem{Cotler:2022fze}
J.~Cotler and S.~Rezchikov,
[arXiv:2202.11737 [hep-th]].

\bibitem{Sonoda:2020vut}
H.~Sonoda and H.~Suzuki,
PTEP \textbf{2021}, no.2, 023B05 (2021)
[arXiv:2012.03568 [hep-th]].


\end{thebibliography}
\end{document}